\documentclass[twocolumn,showpacs,preprintnumbers,amsmath,amssymb]{revtex4}

\usepackage{graphicx}
\usepackage{dcolumn}
\usepackage{bm}

\usepackage{times}

\newcommand\rmd{\mathrm{d}}

\begin{document}
\title{Thermo-magneto coupling in a dipole plasma}
\author{Z. Yoshida, Y. Yano, J. Morikawa, and H. Saitoh}
\affiliation{
Graduate School of Frontier Sciences, The University of Tokyo,
Kashiwa, Chiba 277-8561, Japan.}
\date{\today}

\begin{abstract}
On a dipole plasma, we observe the generation of magnetic moment, 
as the movement of the levitating magnet-plasma compound, in response to 
electron-cyclotron heating and the increase of $\beta$ (magnetically-confined thermal energy).
We formulate a thermodynamic model with interpreting heating 
as injection of microscopic magnetic moment;
the corresponding chemical potential is the ambient magnetic field. 
\end{abstract}


\maketitle

\begin{figure}[bt]
\begin{center}
  \includegraphics[width=0.4\textwidth]{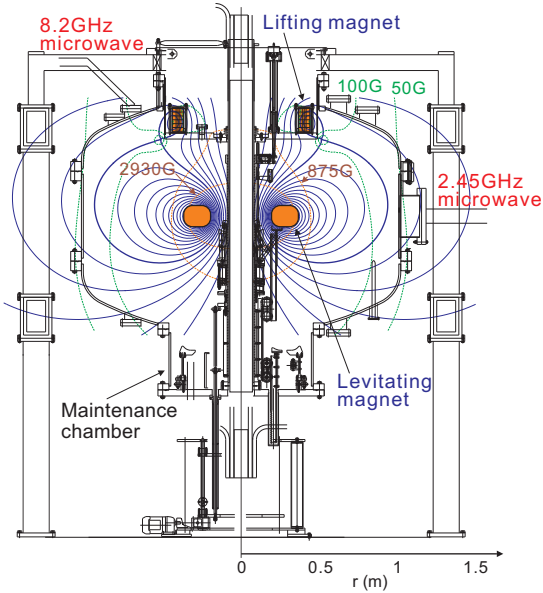}
\end{center}
\caption{ 
Schematic drawing of the RT-1 device.
A dipole magnetic field is produced by the levitating superconducting magnet.
The field strength in the confinement region varies from $0.5$\,T to $0.01$\,T.
Plasma is produced and heated by ECH (8.25 GHz, 25 kW, and 2.45GHz, 20 kW systems). 
}
\label{fig:RT-1}
\end{figure}

The RT-1 device confines a high-temperature
(electron temperature $T_e \sim 10$ keV) 
plasma in a dipole magnetic field that is generated by a levitating superconducting magnet\,\cite{Yoshida2006,Morikawa2007,Saito_PoP,Saito_NF}; see Fig.\,\ref{fig:RT-1}.
When a high-beta (local $\beta \sim 0.7$) plasma is produced,
we observe an appreciable amplitude of vertical motion of the levitating magnet-plasma compound,
while the magnet position is regulated by a feedback control system\,\cite{Yano2010}.
Interpreting this phenomenon form thermodynamic view point, we will
delineate an interesting property of magnetized plasmas.

Let us start by analyzing mechanics.
We denote by $z$ the vertical
displacement of the magnet from the equilibrium position
(we use $r$-$\theta$-$z$ cylindrical coordinates).
From the time-series data of the coil position
and the controlled current ($I_L$) in the lifting coil, 
we can estimate the change of forces acting on the
levitating magnet-plasma compound: denoting by $M$ ($=112$ kg) 
the mass of the magnet (the mass of the plasma is ignorable), 
the equation of motion in the vertical direction can be written as
\begin{equation}
M \frac{\rmd^2 z}{\rmd t^2} = F_m - Mg .
\label{Eq_of_motion}
\end{equation}
On the right-hand side, $Mg$ is the gravity and $F_m$ is the magnetic force in the
vertical direction:
\begin{equation}
F_m := \int \bm{e}_z\cdot(\bm{J}\times\bm{B}_L)\,\rmd^3x
\approx - 2\pi R I B_{L,r} ,
\label{magnetic_moment-1}
\end{equation}
where $\bm{B}_L$ is the magnetic field applied by the lifting magnet
($B_{L,r}$ is its radial component), 
$\bm{J}$ is the current density in the magnet-plasma compound, and
$I$ is the total current in $\theta$-direction.
Here we have approximated $\bm{J}$ by a ring current of radius $R$.
Invoking the conventional magnetic moment ${\cal M}=\pi R^2 I$,
we may write
(using $\nabla\cdot\bm{B}=r^{-1}\partial (rB_{r})/\partial r +\partial B_{z}/\partial z=0$ and
$\partial B_{r}/\partial r \approx B_{r}/r$)
\begin{equation}
F_m = -2 {\cal M} \frac{B_{L,r}}{r}
= {\cal M} \frac{\partial B_{L,z}}{\partial z}.
\label{magnetic_moment-2}
\end{equation}
We define ${\cal M}$ to be positive and $B_{L,r}$ to be negative
($B_{L,z}$ and $\partial B_{L,z}/\partial z$ to be positive), and then,
$F_m$ is positive (upward).
We denote
\begin{equation}
{\cal G} := \frac{\partial B_{L,z}}{\partial z}
\label{define_L}
\end{equation}
to write $F_m = {\cal M}{\cal G}$.
At the equilibrium point ($z=0$), we define ${\cal M}={\cal M}_0$ and
${\cal G}={\cal G}_0$.  The equilibrium condition reads as
${\cal M}_0{\cal G}_0 = Mg$.

While (\ref{magnetic_moment-2}) is derived for the
conventional magnetic moment of a loop current,
we may use it to ``define'' the total magnetic moment of the
magnet-plasma system.
In what follows, we evaluate ${\cal G}$ at the barycenter
of the levitating magnet, and define ${\cal M}:= F_m/{\cal G}$ by the
total magnetic force $F_m$ on the levitating magnet-plasma compound.
Linearizing (\ref{Eq_of_motion}) in the neighborhood of the equilibrium point 
($z=0$, ${\cal G}={\cal G}_0$ and ${\cal M}={\cal M}_0$), we obtain
\begin{equation}
M \frac{\rmd^2 z}{\rmd t^2} = {\cal M}_0\frac{\partial {\cal G}}{\partial z} z
+{\cal M}_0 \frac{\partial {\cal G}}{\partial I_L} \rmd I_L  +{\cal G}_0\rmd{\cal M} ,
\label{Eq_of_motion-linear}
\end{equation}
where $(\partial {\cal G}/\partial I_L)\rmd I_L$ represents the 
variation of ${\cal G}$ due to a
perturbation $\rmd I_L$ in the lifting magnet
(we define the sign of $I_L$ so that $\rmd I_L >0$ increases $B_{L,z}$ and ${\cal G}$)\,\cite{stability}. 
The inertial force on the left-hand-side of (\ref{Eq_of_motion-linear}) 
can be estimated by the time-series data of the coil position. 
Evaluating the first and second terms on the right-hand side of (\ref{Eq_of_motion-linear}) by
measured $\rmd z\,(=z)$ and $\rmd I_L$, we obtain
the remaining third term, by which we can derive $\rmd{\cal M}$.

\begin{figure}[bt]
\begin{center}
  \includegraphics[width=0.4\textwidth]{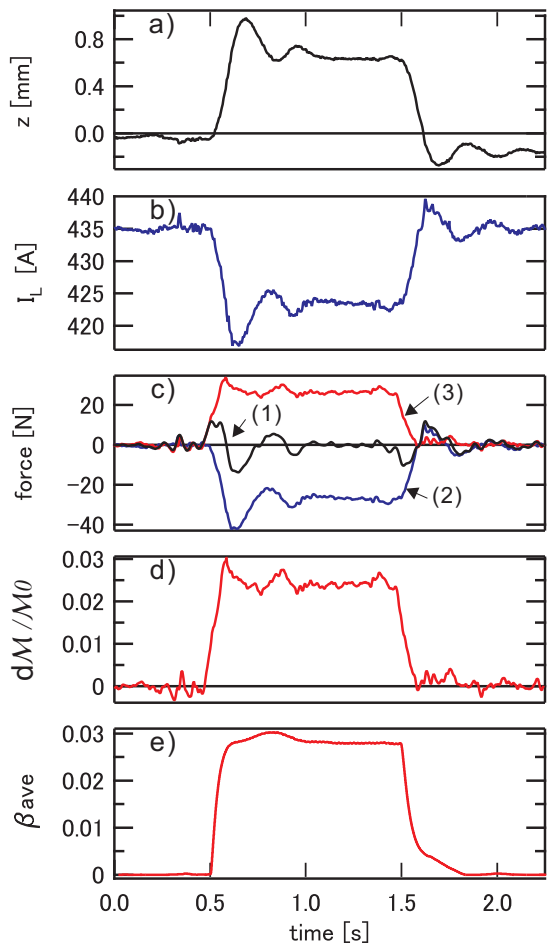}
\end{center}
\caption{ 
Typical waveforms of 
(a) the vertical plasma position $z$ (measured by laser position sensors),
(b) the lifting-magnet current $I_L$ (feedback controlled in response to the position signal),
(c) (1) the inertial force, (2) magnetic force by perturbed $\rmd I_L$, and (3) the
remaining term of (\ref{Eq_of_motion-linear}) corresponding to the magnetic force by the perturbed magnetic moment $\rmd{\cal M}$,
(d) the perturbed magnetic moment normalized by the magnetic moment ${\cal M}_0$ 
of the superconducting magnet, and 
(e) the normalized plasma energy ($\beta$ averaged over the plasma volume) estimated by diamagnetic signals. ECH is injected for $0.5 < t < 1.5$\,s.
}
\label{fig:waveform}
\end{figure}

\begin{figure}[bt]
\begin{center}
  \includegraphics[width=0.4\textwidth]{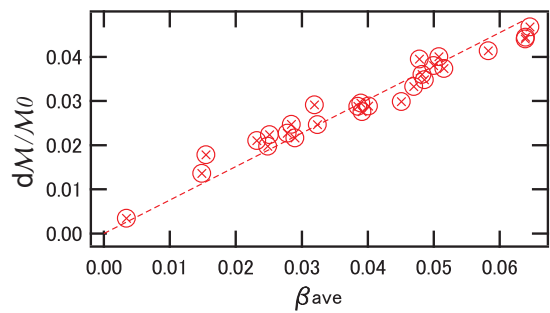}
\end{center}
\caption{
Experimental relation between the plasma-volume-averaged $\beta_{\rm ave}$ 
(estimated by the diamagnetic signal) and the increment of the magnetic moment $\rmd{\cal M}$
(normalized by the magnetic moment ${\cal M}_0$ of the superconducting magnet).
The experimental scaling (dotted line) shows $\rmd{\cal M}/{\cal M}_0 = 0.75\beta_{\rm ave}$
}
\label{fig:scaling}
\end{figure}

We observe that the magnetic moment ${\cal M}$ increases as the plasma is heated;
in Fig.\,\ref{fig:waveform} we compare the waveforms of the change in ${\cal M}$ and
the volume-average $\beta$ estimated by diamagnetic signals
(for the detail of the measurement, see\,\cite{Saito_NF}).
We may explain the increased magnetic moment in terms of the
\emph{diamagnetic current} driven by the plasma pressure.
The unique structure of this device ---a ``levitating'' confinement system---
therefore provides us with a particular method of estimating the 
plasma pressure by measuring the mechanical motion of the magnet;
in Fig.\,\ref{fig:scaling} we show an experimental
relation between $\beta$ and $\rmd{\cal M}$.

The main theme of this brief communication, however, is not the practical application 
of the magnetic moment for measurement.
Examining this phenomenon from a thermodynamic view point, we notice an
interesting implication, and that is the subject of present practice.
Injecting electron cyclotron heating (ECH) power, we 
increase the internal (thermal) energy of the plasma (electrons).
In the language of thermodynamics,
giving a heat $\delta Q$ causes a change $\rmd U$ of the internal energy $U$
(here we denote a general variation by $\delta X$,
while a variation of a state variable $Y$ is written as $\rmd Y$);
the energy $U$ is the combination of the thermal, mechanical, gravitational and electromagnetic energies,
and $\delta Q$, in general, may cause variations in every component of the
energy, resulting in changes in macroscopic quantities including those
mechanical (vertical velocity), gravitational (vertical position), and electromagnetic
(magnetic moment). 
Writing the first law as
\begin{equation}
\rmd U = \delta W + \delta Q ,
\label{first_law}
\end{equation}
the term $\delta W$ represents whole such contributions from macroscopic quantities to the energy balance.
In textbook thermodynamics, we often assume that $\delta W= -P\rmd V$
with a pressure $P$ and volume $V$, and then, the coupling of the
thermodynamic energy and the macroscopic mechanical energy is only through
compressible motion of fluid.
Needless to say, possible processes are much more rich in a plasma.

As mentioned above, we observe that heating $\delta Q$ causes a change in
the magnetic moment ${\cal M}$ and subsequent changes in the vertical position and
(feedback controlled) lifting-magnet current.
To describe the ``thermodynamics'' of this system, 
we have to formulate the relations among $\delta Q$, $\rmd{\cal M}$, $\rmd I_L$, and $\rmd z$.
Here we proffer a ``grand-canonical model'' to understand this thermo-magneto coupling.
We do not intend to challenge the aforementioned elementary understanding
in terms of the diamagnetic current.
Instead, our new perspective will delineate an interesting property of a
magnetized plasma in a more succinct picture.

The energy of a magnetic moment\,\cite{remark:magnetic_energy} is at the
core of the first law connecting the plasma, the magnet, the heating system, and the
lifting system. 
When an external magnetic field $\bm{B}_L$ is applied, the magnetic moment ${\cal M}$ has a
\emph{mechanical potential energy}\,\cite{magnetic_moment}
\begin{equation}
{V}_m = -{\cal M}{\cal B}_L  ,
\label{energy_of_m}
\end{equation}
where ${\cal B}_L$ is the average of $\bm{e}_z\cdot\bm{B}_{L}$ over the levitating
magnet-plasma compound.
As well known\,\cite{Feynmam}, the ``total energy'' of a magnetic moment, including the
electric energies of the levitating and lifting currents, is $-{V}_m$,
but we must use ${V}_m$ to derive mechanical forces and corresponding works.
Combining ${V}_m$ with the gravitational energy $U_g=Mg z$
and the kinetic energy $U_k = p_z^2/(2M)$
($p_z = M\rmd z/\rmd t$ is the momentum), we obtain a \emph{Hamiltonian}
\begin{equation}
H := U_k + U_g + {V}_m.
\label{Hamiltonian}
\end{equation}
The corresponding Hamilton's equation of motion
reproduces (\ref{Eq_of_motion-linear}).
The explicit dependences of ${V}_m$ on the 
parameters $I_L$ and ${\cal M}$ yield changes of $H$:
\begin{equation}
\frac{\rmd H}{\rmd t} = \frac{\partial H}{\partial t}
= \frac{\partial {V}_m}{\partial I_L}\frac{\rmd I_L}{\rmd t}
+ \frac{\partial {V}_m}{\partial {\cal M}}\frac{\rmd {\cal M}}{\rmd t}.
\label{Hamiltonian-dt}
\end{equation}

As already remarked, ${V}_m$
is not the right energy to be inserted into the first law; 
we have to add the electric energies in the levitating and lifting systems,
which amounts $2 {\cal M}{\cal B}_L= -2 {V}_m$\,\cite{Feynmam}.
Hence, the magnetic moment acquires an energy, when put in an external
magnetic field $\bm{B}_L$,
\begin{equation}
U_m = {V}_m + 2{\cal M}{\cal B}_L
= {\cal M}{\cal B}_L = - {V}_m.
\label{U_of_magnetic_moment}
\end{equation}
Notice the flip of the sign of energy.
In addition to this \emph{mutual energy},
the magnetic field of the total system 
(consisting of $\bm{B}_{D}$
that is produced by the dipole magnet-plasma compound
and $\bm{B}_{L}$ that is produced by the lifting magnet)
has also the \emph{self-energy} $U_s$
that may be written as\,\cite{remark:magnetic_energy}
\begin{eqnarray}
U_s &=& \frac{1}{2\mu_0} \int \left( B_D^2 + B_L^2 \right) \rmd^3 x
\nonumber \\
&=& \frac{1}{2}{{\cal B}_{D}}{\cal M} + \frac{1}{2\mu_0}\int B_{L}^2\,\rmd^3 x,
\label{self_magnetic_energy}
\end{eqnarray}
where ${\cal B}_{D}$ is an average of $\bm{e}_z\cdot\bm{B}_{D}$.

Including the thermal energy $U_t$ of the plasma, the total energy of the system is
\begin{equation}
U= U_k+U_g+U_m + U_s + U_t .
\label{grand_canonical}
\end{equation}
The first law, combined with the ``mechanical law'' (\ref{Hamiltonian-dt}),
reads as
\begin{equation}
\rmd U = {\cal M}\rmd {\cal B}_{L} + {\cal B}_{L} \rmd{\cal M} + \rmd U_s + \rmd U_t .
\label{dU}
\end{equation}
We find that ${\cal M}\rmd {\cal B}_{L}={\cal M}(\rmd{\cal B}_{L}/\rmd I_L)\rmd I_L$
contributes to the mechanical work $\rmd H$ (on the magnet-plasma subsystem)
by $(\partial V_m/\partial I_L)\rmd I_L
= -[{\cal M}(\partial^2{\cal B}_{L}/\partial I_L\partial z)\rmd I_L] \rmd z$
(notice the flip of the sign).
On the other hand, $\rmd{\cal M}$ is ``caused'' by heating $\delta Q$, thus
we may relate the term ${\cal B}_{L} \rmd{\cal M}$ with $\delta Q$
(the latter also includes energy loss).

To delineate the relation between $\rmd{\cal M}$ and $\delta Q$,
we invoke the microscopic magnetic moment $\mu = (m_e v_c^2)/(2B)$,
where $B$ is the local magnetic field in the plasma region,
$m_e$ is the mass of an electron, and $v_c$ is the velocity of cyclotron motion.
The power of ECH, first of all, increases $v_c^2$
(and then, excites macroscopic processes).
The perpendicular thermal energy $U_{t,\perp}$ is the sum of $B\mu_j$ 
over all particles (labeled by $j=1,2,\cdots$).
With an average magnetic field ${\cal B}$, we write
\begin{equation}
U_{t,\perp} := \sum_j B \mu_j = {\cal B} \sum_j \mu_j .
\label{chemical_potential}
\end{equation}
In view of (\ref{chemical_potential}),
we may rephrase ``heating'' as injection of microscopic magnetic moments $\mu_j$,
and then, ${\cal B}$ is an effective \emph{chemical potential}. 

To relate the microscopic magnetic moments $\mu_j$ with the macroscopic one
${\cal M}_p$ (we denote by ${\cal M}_p$ the plasma's contribution to ${\cal M}$), we put
\begin{equation}
{\cal M}_p = {\cal D} \sum_j \mu_j
\label{chemical_potential2}
\end{equation}
with a geometric factor ${\cal D}$, which we can estimate as follows.
By the levitating magnets's current $I_0$ and
the length scale $\ell$ of poloidal magnetic field lines
($\ell \sim 2\pi a$ with a minor radius $a$), we estimate
${\cal B}=\mu_0 I_0/\ell$.
Normalizing by ${\cal M}_0 = \pi R^2 I_0$,
we obtain
\begin{equation}
\frac{{\cal M}_p}{{\cal M}_0} = {\cal D} \frac{U_{t,\perp}}{\pi R^2\ell{\cal B}^2/\mu_0}
=  {\cal D} \frac{V}{2\pi R^2\ell} \beta_\perp ,
\label{scaling}
\end{equation}
where $V$ is the volume of the plasma and 
$\beta_\perp := U_{t,\perp}/(V{\cal B}^2/2\mu_0)$ is the average beta ratio of the perpendicular
plasma pressure.
On the other hand, we estimate ${\cal M}_p=\pi R^2 I_p$, where $I_p$ is the diamagnetic current
induced by the perpendicular pressure $P_\perp$.  Estimating
$I_p= \ell P_\perp/{\cal B}$, we obtain
\begin{equation}
\frac{{\cal M}_p}{{\cal M}_0} = \frac{ \beta_\perp}{2} .
\label{scaling2}
\end{equation}
Figure\,\ref{fig:scaling} shows a reasonable agreement.
Comparing (\ref{scaling}) and (\ref{scaling2}), we estimate 
${\cal D}= \pi R^2\ell/V\sim R/a$.
Since the change of the superconductor's current
in response to $\rmd z$, $\rmd I_L$ or $\rmd{\cal M}_p$ is of second order,
we may assume $\rmd{\cal M}=\rmd{\cal M}_p$.

Now we have a more explicit representation of the thermo-magneto coupling processes
included in the first law (\ref{dU}):
denoting by $U_{t,\parallel}$ the remaining parallel component of the thermal energy $U_t$
and by ${\cal M}_L$ the coefficient such that
$\int B_{L}^2\,\rmd^3 x/\mu_0={\cal M}_L{\cal B}_{L}$,
\begin{eqnarray}
\rmd U &=& \left({\cal M}+\frac{{\cal M}_L}{2}\right) \frac{\rmd {\cal B}_{L}}{\rmd I_L}\rmd I_L 
\nonumber \\
& & +
\left({\cal B}_{L} + \frac{{\cal B}_{D}}{2} +\frac{{\cal B}}{{\cal D}}\right)\rmd{\cal M}
\nonumber \\
& & + \rmd U_{t,\parallel} + \delta Q'.
\label{dU2}
\end{eqnarray}
The first term on the right-hand side (induced by $\rmd I_L$) is the process connected to the lifting magnet system.
The second term (induced by $\rmd{\cal M}$) is the ``ECH heating'' $\delta Q_{ECH}$
(or, in our language, \emph{injection of magnetic moments});
the component $({\cal B}/{\cal D})\rmd{\cal M}$ goes to the thermal energy $U_{t,\perp}$,
while the other components change macroscopic magnetic energies $U_m$ and $U_s$,
as well as mechanical energies $U_k$ and $U_g$ (through the mechanical potential energy $V_m=-U_m$),
which we observe as the change of $z$.
The remaining abstract terms $\rmd U_{t,\parallel}$ (parallel energy change) and $\delta Q'$
(heat processes including thermal conduction, energy loss with particle transport, etc.) 
are not the direct subject of the present analysis.


We have made an attempt to understand and interpret the observed 
macroscopic thermo-magneto coupling in a dipole plasma
produced on the RT-1 magnetospheric device.
The most abstract thermodynamic first law (\ref{first_law})
has been given a more concrete and dissected form (\ref{dU2}) that elucidates the
internal and external thermo-magneto processes;
the conventional expression of ECH as heating $\delta Q$ has been 
rewritten an injection of magnetic moment $\rmd{\cal M}$,
and its partition into different terms of energy has been specified.

What is rather nontrivial is that a magnetic moment
$\bm{m}$ is an axial vector (or, a pseudo-vector) having an odd parity;
the ${\cal M}$ is the $z$-component of $\bm{m}$
(i.e. $\bm{m}={\cal M}\bm{e}_z$),
which can be regarded as a pseudo-scalar.
Multiplying $\bm{m}$ (${\cal M}$) by the other axial vector $\bm{B}$ 
(a pseudo-scalar ${\cal B}$), we obtain a scalar that can be related to an energy or
some thermodynamic potential.
Remember that the enthalpy $U+PV$ of a neutral fluid couples with 
a product $\nabla P\cdot\bm{u}$ of two vectors $\nabla P$ and $\bm{u}$ 
(fluid velocity).
Or, more simply, we write the work as $P\rmd V$ (or $-V\rmd P$ for estimating enthalpy)
with two scalars $P$ and $V$.
Relating the pressure $P$ to the thermal energy by an
equation of state, we can close a thermodynamic relation.
To describe a thermodynamic model of a plasma, therefore, we have to
find a relation between an axial vector (pseudo-scalar) and the thermal energy
---there must be an intrinsic mirror-symmetry breaking to make such a relation possible.
We have proposed a ``grand-canonical model'' with a pseudo-scaler chemical potential
(that is the ambient magnetic field introducing the symmetry breaking).

We end this brief communication with a comment to extend the scope of the
paradigm of pseudo-scalar chemical potentials; 
different mechanisms of magnetic field (axial vector) generation
can be related on a unified perspective.
Remember that the \emph{helicity} $K:=\int\bm{A}\cdot\bm{B}\,\rmd^3x$
is also a pseudo-scalar, which measures the twist, linking, and writhe of magnetic field lines\,\cite{Moffatt}.
A ``helicity injection'' into some thermodynamic (or turbulent) system
may create a current with twisting magnetic field lines.
This idea has been successfully demonstrated in plasma experiments\,\cite{Schoenberg1984,Schoenberg1988,Ono}.
In this case, we invoke a pseudo-scalar coefficient $\lambda$ 
and define a magnetohydrodynamic free energy as
$F:=\int B^2\,\rmd^3x/(2\mu_0) - \lambda K$.
The minimizer of $F$ gives an equilibrium
magnetic field with a finite current (in this case, $\bm{J}$ parallels $\bm{B}$)\,\cite{JBT}.
The $\lambda$ (called Beltrami-parameter) can be interpreted as a pseudo-scalar chemical potential,
and, introducing a grand-canonical ensemble of magnetic and flow fields, 
a Boltzmann distribution with a finite helicity can be formulated\,\cite{Ito}.

\acknowledgments
We acknowledge the support given by the RT-1 project members.
This work was supported by the Grant-in-Aid for Scientific Research
No. 23224014 from 
Japanese Ministry of Education, Science and Culture.



\begin{thebibliography}{}
\bibitem{remark:magnetic_energy}
Because the levitating magnet-plasma system consists of 
a superconductor and high-temperature plasma,
the magnetic flux is conserved.  Under the
flux-conserving condition, the magnetic energy is better evaluated in terms of magnetic moment.
The most general expression of the magnetic energy is 
(ignoring the displacement current)
$U_M = \int B^2\,\rmd^3 x/(2\mu_0) =  \int \bm{A}\cdot\bm{J} \,\rmd^3 x/2$.
In an axisymmetric dipole configuration, $\bm{J}=J_\theta\bm{e}_\theta$, and then,
$U_M = \pi \int \psi J_\theta\,\rmd r\rmd z $,
where $\psi := rA_\theta=\int B_z r\rmd r$ is the magnetic flux.
With an average value $\overline{\psi}=\overline{B_z}S/(2\pi)$
(at $r=R$, $\psi=\overline{\psi}$; $S:=2\pi\int r\rmd r$ is the area of the disk of radius $R$,
and $\overline{B_z}$ is the average of $B_z$ on the disk), 
and the current $I=\int J_\theta\,\rmd r\rmd z$, we may write
$U_M = \overline{B_z} S I/2 = \overline{B_z} {\cal M}/2$.
Decomposing $\bm{B}=\bm{B}_D + \bm{B}_L$ ($\bm{B}_{D}$ is the dipole magnetic field;
$\nabla\times\bm{B}_{D}=\mu_0\bm{J}$ produces ${\cal M}$ ), we may write 
$U_M= \int \bm{A}_{D}\cdot\bm{J} \,\rmd^3 x/2+\int \bm{A}_{L}\cdot\bm{J} \,\rmd^3 x + \int B_{L}^2\,\rmd^3 x/(2\mu_0)
= \overline{B_{S,z}}{\cal M}/2 + \overline{B_{L,z}}{\cal M} + \int B_{L}^2\,\rmd^3 x/(2\mu_0)$.
The second term is the ``mutual'' energy $U_m$.
\bibitem{magnetic_moment}
In general, a magnetic moment is an axial vector $\bm{m}$,
and ${V}_m = -\bm{m}\cdot\bm{B}$.
Here, ${\cal M}$ is the $z$-component (i.e. $\bm{m}={\cal M}\bm{e}_z$).
The axisymmetric geometry of the present system
allows us to omit the torque on the magnetic moment.
\end{thebibliography}
\end{document}